%% file: main.tex
\documentclass{article}
\usepackage[a4paper,total={6in, 8in}]{geometry}
\usepackage[utf8]{inputenc}
\usepackage[T1]{fontenc}
\usepackage{amsfonts}
\usepackage{nicefrac}
\usepackage{microtype}
\usepackage{braket}
\usepackage{graphicx}
\usepackage{doi}
\usepackage{amsmath,amsfonts,amsthm,amssymb}
\usepackage[usenames,dvipsnames]{xcolor}
\usepackage{hyperref}
\usepackage{xcolor}
\usepackage{bbold}

\definecolor{linkcolour}{rgb}{0,0.2,0.6}
\definecolor{citecolour}{rgb}{0,0.5,0.2}
\hypersetup{colorlinks,breaklinks,urlcolor=linkcolour,linkcolor=linkcolour,citecolor=citecolour}
\usepackage[style=numeric,sorting=none,subentry]{biblatex}
\DeclareMathOperator{\sign}{sign}

\newcommand{\de}{\mathrm{d}}

\title{Quantum backreaction for overspinning BTZ geometries}
\author{Olaf Baake$^{2,1}$ \footnote{olaf.baake@gmail.com}
$\,$ and Jorge Zanelli$^{1,3}$ \footnote{jorge.zanelli@uss.cl}
\\[12pt]
$^1$\textit{Centro de Estudios Cient\'{\i}ficos (CECs), Arturo Prat 514, Valdivia, Chile} \\
$^2$\textit{Instituto de Matem\'aticas,	Universidad de Talca, Casilla 747, Talca 3460000, Chile}\\
$^3$\textit{Universidad San Sebasti\'an, General Lagos 1163, Valdivia, Chile}\\
}

\begin{document}
\maketitle
\begin{abstract}
We examine the semiclassical backreaction of a conformally coupled scalar field on an overspinning BTZ geometry. This extends the work done on a similar problem for ($2+1$)- AdS geometries of the BTZ family with $|M|>|J|$. The overspinning classical solutions corresponds to $|M|<|J|$  and possess a naked singularity at $r=0$. Using the renormalized quantum stress-energy tensor for a conformally coupled scalar field on such a spacetime, we obtain the semiclassical Einstein equations, which we attempt to solve perturbatively. We show that the stress-energy tensor is non-renormalizable in this approach, and consequently the perturbative solution to the semiclassical equations in the overspinning case does not exist. This could be an indication of the fact that the naked singularity at the center of an overspinning geometry is of a more severe nature than the conical singularity found in the same family of BTZ geometries.
\end{abstract}

\input{introduction}
\input{overspinning_btz}
\input{rset}

\input{metric}
\input{conclusions}


\section*{Acknowledgements}
\label{sec:ack}
We thank C. Mart\'inez, M. Hassa\"{i}ne and Steen Ryom-Hansen for many enlightening discussions. OB is funded by the PhD scholarship of the University of Talca. This work has been partially funded by grant N$^o$ 1220862 from ANID/Fondecyt.

\printbibliography
\end{document}

%% file: introduction.tex
\section{Introduction}
\label{sec:intro}
Since the dawn of general relativity, many black hole solutions to Einstein's field equations have been found. All these black holes contain a spacetime singularity hidden by an event horizon. However, for some range of values of the integration constants (mass $M$, angular momentum $J$, electric charge $Q$) these solutions have  no event horizon. Although  paradoxical, these naked singularities are exact solutions to the classical equations of general relativity as well. In the vicinity of a naked singularity causality and other physical laws can be arbitrarily violated, which is why Roger Penrose suggested the existence of a (weak) cosmic censorship  principle in nature \cite{1969NCimR...1..252P}, requiring singularities to be hidden behind an event horizon. In that case, an outside observer would be causally disconnection from the singularity.  

Classically, naked singularities cannot be ruled out on mathematical grounds, and it is difficult to prove that every possible collapse process leads to the formation of an event horizon. The fact that so far no naked singularities have been observed in the universe may be interpreted as an indication that, in the strong gravity regime near a singularity, quantum gravity effects dominate eliminating singularities altogether, or at least making sure that a horizon forms around them.

The accumulation of experiments and observations that confirm the predictions of general relativity puts very tight constraints on possible theories incorporating both general relativity and quantum theory. Since both theories are so well established in their regimes, it is sensible to look for a common area where a semi-classical approach could be used to obtain a better understanding of the issues at hand. Calculating quantum effects on a curved background spacetime is notoriously difficult, but in (2+1)-dimensional AdS spacetime this problem becomes significantly simpler and still provide meaningful information to learn from. 

The  Ba\~nados-Teitelboim-Zanelli (BTZ) black hole in (2+1)-dimensional AdS spacetime \cite{Banados:1992wn, Banados:1992gq}, obtained for $M\geq |J|$ are particularly interesting geometries in this respect, but these are not the only solutions of physical interest in this theory and with the same global symmetries. Locally constant curvature 2+1 spacetimes include, besides the BTZ black hole family, the self-dual Coussaert-Henneaux spacetimes \cite{Coussaert:1994tu}, and the toroidal time-dependent geometries \cite{Ayon-Beato:2004ehj}, with global isometry groups $SO(2)\times \mathbb{R}$ $SO(2)\times SO(2,1)$ and $SO(2)\times SO(2)$, respectively.

Recently, the quantum back reaction on the classical singularities was studied for several geometries, including static, rotating and extremal BTZ black holes, as well as for static and rotating conical naked singularities \cite{Casals:2016ioo, Casals:2016odj, Casals:2018ohr, Casals:2019jfo}. The naked singularities considered in these papers are continuations of the BTZ spacetime to the case of negative mass \cite{Miskovic:2009}. The interesting aspect of this result is that the quantum fluctuations of a conformally coupled scalar field generate a non-vanishing stress energy-momentum tensor that through Einstein's equations produces aback-reacted geometry with a horizon of order Planck length in radius. This dressing up of the naked singularity, turning it into a black hole, could be viewed as a mechanism that implements cosmic censorship. These results have also been confirmed by an alternative holographic approach in \cite{Emparan:2020znc}.

Here we are concerned with the overspinning BTZ spacetime, which occurs if the absolute value of the angular momentum is greater than that of the mass. This geometry is also endowed with a naked singularity at $r=0$, as in the case of the conical singularity obtained for $M\leq - |J|$.

We show that the stress-energy tensor contains incurable divergences, making the perturbative ansatz to the semiclassical equations of motion ill-defined. While the equations of motion can still be formally integrated, the first order corrections to the metric functions would become large, further demonstrating the inapplicability of a perturbative approach to this type of geometry. This strongly suggests that the naked singularity of an overspinning geometry is of a more severe nature than the conical singularities appearing in the other BTZ geometries so that they cannot be cured by a perturbative quantum censor.

%% file: overspinning_btz.tex
\section{Overspinning BTZ space-time}
\label{sec:geo}
The rotating BTZ metric \cite{Banados:1992wn,Banados:1992gq}, is given by
\begin{equation} \label{BTZ-metric}
    \de s^2 = -\left( \frac{r^2}{l^2} - M \right) \de t^2 - J \de t \de \theta + \left( \frac{r^2}{l^2} - M + \frac{J^2}{4 r^2} \right)^{-1} \de r^2 + r^2 \de \theta^2,
\end{equation}
where the coordinate ranges are: $-\infty < t < \infty$, $0 < r < \infty$ and $0 \leq \theta < 2 \pi$, $\Lambda=-l^{-2}$ is the cosmological constant, and $M$ and $J$ are mass and angular momentum respectively. This metric describes different spacetimes that can be classified by the values of $M$ and $J$ which determine the nature of the four roots of the equation $g^{rr}=0$,
\begin{equation}
    \lambda_{\pm} = \frac{l}{2} \left[ \sqrt{M + \frac{J}{l}} \pm \sqrt{M - \frac{J}{l}} \right]\,. 
\end{equation}
These roots are real for $M\geq |J|/l$ (black holes) and take complex values for $M<|J|/l$ (naked singularities). The full classification is explained in detail in \cite{Banados:1992gq}, but here we will consider the so-called overspinning geometry ($|M|l < |J|$). This geometry was examined in \cite{Briceno:2021dpi} through the study of classical geodesics around it. In particular, we will analyze the back reaction of the geometry to the presence of a conformally coupled quantum scalar field, following the steps in \cite{Casals:2016ioo,Casals:2016odj,Casals:2018ohr,Casals:2019jfo}, where the back reaction for conical naked singularities in the parameter range $M\leq -|J|$ was  studied. 

The starting point of the analysis is the observation that the BTZ spacetimes \eqref{BTZ-metric} are quotients of the universal covering of anti-de Sitter space-time ($\mathrm{CAdS}_3$) by an appropriate Killing vector field \cite{Banados:1992gq}. The constant negative curvature spacetime $\mathrm{AdS}_3$ is defined by a pseudosphere of radius $l$ embedded in $\mathbb{R}^{(2,2)}$ as
\begin{align}
    \eta_{AB} X^A X^B &= - \left( X^0 \right)^2 + \left( X^1 \right)^2 + \left( X^2 \right)^2 - \left( X^3 \right)^2 = -l^2\,.
\end{align}
The metric reads
\begin{align}
    \eta_{AB} dX^A dX^B &= - \left( dX^0 \right)^2 + \left( dX^1 \right)^2 + \left( dX^2 \right)^2 - \left( dX^3 \right)^2,
\end{align}
where the embedding coordinates $X^A$ must be specified as functions of $(t,r,\theta)$. As shown in \cite{Briceno:2021dpi}, the overspinning geometry \eqref{BTZ-metric} with $|M|<|J|$ corresponds to embedding coordinates given by
\begin{align} \nonumber
    X^0 =& \frac{l}{2} \sqrt{A+1} \cosh \left[a \left(t/l-\theta \right)\right] \left\{\cos \left[b \left(\theta +t/l\right)\right] - \sin \left[b \left(\theta +t/l\right)\right]\right\} \\
    +& \epsilon \frac{l}{2} \sqrt{A-1} \sinh \left[a \left(t/l-\theta \right)\right] \left\{\sin \left[b \left(\theta +t/l\right)\right] + \cos \left[b \left(\theta +t/l\right)\right]\right\},\\ \nonumber
    X^1 =& \frac{l}{2} \sqrt{A+1} \sinh \left[a \left(t/l-\theta \right)\right] \left\{\cos \left[b \left(\theta +t/l\right)\right] - \sin \left[b \left(\theta +t/l\right)\right]\right\} \\
    +& \epsilon \frac{l}{2} \sqrt{A-1} \cosh \left[a \left(t/l-\theta \right)\right] \left\{\sin \left[b \left(\theta +t/l\right)\right] + \cos \left[b \left(\theta +t/l\right)\right]\right\},\\ \nonumber
    X^2 =& \frac{l}{2} \sqrt{A+1} \sinh \left[a \left(t/l-\theta \right)\right] \left\{\sin \left[b \left(\theta +t/l\right)\right] + \cos \left[b \left(\theta +t/l\right)\right]\right\} \\
    -& \epsilon \frac{l}{2} \sqrt{A-1} \cosh \left[a \left(t/l-\theta \right)\right] \left\{\cos \left[b \left(\theta +t/l\right)\right] - \sin \left[b \left(\theta +t/l\right)\right]\right\},\\ \nonumber
    X^3 =& \frac{l}{2} \sqrt{A+1} \cosh \left[a \left(t/l-\theta \right)\right] \left\{\sin \left[b \left(\theta +t/l\right)\right] + \cos \left[b \left(\theta +t/l\right)\right]\right\} \\
    -& \epsilon \frac{l}{2} \sqrt{A-1} \sinh \left[a \left(t/l-\theta \right)\right] \left\{\cos \left[b \left(\theta +t/l\right)\right] - \sin \left[b \left(\theta +t/l\right)\right]\right\},
\end{align}
where
\begin{equation}
a = \frac{\sqrt{|J|/l+M}}{2}, \qquad b = \frac{\sqrt{|J|/l-M}}{2},
\qquad A = \frac{2\sqrt{\frac{J^2}{4} + \frac{r^4}{l^2} - M r^2}}{\sqrt{J^2 - l^2 M^2}},
\end{equation}
with $\epsilon = \sign(M-r^2/l^2)$. Note that both cases ($\epsilon = \pm 1$) lead to the same RSET, and hence to the same end results.\footnote{Without loss of generality, we will assume $J>0$ for the rest of this work.} 

The overspinning BTZ space-time is now obtained through identifications generated by a Killing field $\xi$, which in this case given by \cite{Banados:1992gq,Briceno:2021dpi}
\begin{align}
\label{eq:killing_vector}
   \xi &= -a( J_{01} - J_{23} ) + b( J_{03} - J_{12} ),
\end{align}
which can be written as $\xi = \frac{1}{2}\omega^{AB} J_{AB}$, where the antisymmetric matrix $\omega^{AB}$ characterizes the identification. The Killing field in matrix form reads
\begin{align}
    \xi =& \begin{pmatrix}
 0 & -a & 0 & -b \\
 -a & 0 & -b & 0 \\
 0 & b & 0 & -a \\
 b & 0 & -a & 0 \\
\end{pmatrix}.
\end{align}

The identification in the embedding space $\mathbb{R}^{(2,2)}$ under the action of the Killing field is a mapping defined by the matrix, $H(\xi) = e^{2 \pi \xi}$, which takes the form
\begin{align}
    H =& \begin{pmatrix}
 C(a) c(b) & -S(a) c(b) &  S(a) s(b) & -C(a) s(b) \\
-S(a) c(b) &  C(a) c(b) & -C(a) s(b) &  S(a) s(b) \\
-S(a) s(b) &  C(a) s(b) &  C(a) c(b) & -S(a) c(b) \\
 C(a) s(b) & -S(a) s(b) & -S(a) c(b) &  C(a) c(b) \\
\end{pmatrix}\, ,
\end{align}
where $C(a)\equiv \cosh(2\pi a)$, $S(a)\equiv \sinh(2\pi a)$ $c(b) \equiv \cos(2\pi b)$, and $s(b)\equiv \sin(2\pi b)$. 

An important feature of the Killing vector \eqref{eq:killing_vector} is that the boost and rotation generators $K\equiv J_{01} - J_{23}$ and $J\equiv J_{03} - J_{12}$ commute, $[K,J]=0$. Consequently, $H=e^{2\pi \xi}$ can be factored as $H = H_a \cdot H_b = H_b \cdot H_a$, where $H_a= H|_{b=0}$ and $H_b = H|_{a=0}$. Iterating the identification by $H$ is equivalent to acting with

\begin{equation} 
H^n = \begin{pmatrix}
 C(n a) c(n b) & -S(n a) c(n b) &  S(n a) s(n b) & -C(n a) s(n b) \\
-S(n a) c(n b) &  C(n a) c(n b) & -C(n a) s(n b) &  S(n a) s(n b)  \\
-S(n a) s(n b) &  C(n a) s(n b) &  C(n a) c(n b) & -S(n a) c(n b) \\
 C(n a) s(n b) & -S(n a) s(n b) & -S(n a) c(n b) &  C(n a) c(n b) 
\end{pmatrix}\,
= H_a^n \cdot H_b^n\;.
\end{equation}

Quotienting a manifold by a rotation Killing vector requires the identification angle to be a rational fraction of $2\pi$. Otherwise, each point is identified with infinitely many images which densely cover a circle, and the resulting image set would not be a smooth manifold \cite{Casals:2019jfo}. This means that the coefficient $b$ in \eqref{eq:killing_vector} must be rational, namely, 
\begin{equation}\label{eq:b-rational}
    b=k/m,
\end{equation}
with $k,m$ relative primes. No restrictions are necessary for $a$, as boosts act transitively in a non-compact manner. Note that the $m$-th iteration produces a pure boost (and a rotation by $2k\pi$, which is equivalent to the identity, $H_b^m=\mathbb{1}$). In fact, we can treat the rotated plane and the boosted plane separately by splitting the identification matrix as follows: consider writing $n = q m + p$, where $p \in \{ 0, 1, \dots, m-1 \}$, $q \in \{0, 1, \dots, \infty \}$ and $m$ is some positive integer.

Hence, the powers of $H=H_a \cdot H_b$ can be arranged as follows
\begin{align}
\label{eq:sum_structure}
\begin{matrix}
 \mathbb{1} & H_a H_b & H_a^2H_b^2 & H_a^3H_b^3 & \dots & H_a^{m-1}H_b^{m-1} \\
 H_a^m & H_a^{m+1} H_b & H_a^{m+2}H_b^2 & H_a^{m+3}H_b^3 & \dots & H_a^{2m-1} H_b^{m-1} \\
 H_a^{2m} & H_a^{2m+1} H_b & H_a^{2m+2}H_b^2 & H_a^{2m+3} H_b^3 & \dots & H_a^{3m-1} H_b^{m-1} \\
\vdots & \vdots & \vdots & \vdots & \vdots & \vdots
    \end{matrix} \;.
\end{align}
Here each column corresponds to a fixed $p$ and includes infinitely many boosts, while each row has a fixed $q$ comprising a finite set of  rotations. In this pattern, an interesting observation becomes apparent. First note that $H_a$ is precisely the identification matrix of the rotating non-extremal BTZ black hole, and $H_b$ the identification matrix of the rotating non-extremal naked singularity \cite{Casals:2019jfo}. Now, using trigonometric identities, one can write in general, as can be seen in \eqref{eq:sum_structure},
\begin{align} \label{splitting}
    H^{qm+p} = H_a^{qm} H_a^p H_b^p = H_{a \cdot m}^{q} H_a^p H_b^p,
\end{align}
so that the $p$-th column reads
\begin{equation}
    H_a^p H_b^p \left\{ \mathbb{1}, H_{a \cdot m}^{1}, H_{a \cdot m}^{2}, H_{a \cdot m}^{3}, \cdots \right\}.
\end{equation}
Or in other words, each column contains the powers of the identification matrix associated with the rotating non-extremal black hole, multiplied by some constant. 

%% file: rset.tex
\section{Renormalized stress tensor}
\label{subsec:rset}
To describe the quantum effects on the spacetime geometry, in particular the backreaction of the naked singularity to the presence of a quantum field, we consider the semi-classical Einstein equations
\begin{equation}
\label{eq:einstein-eqs}
    G_{\mu\nu} - l^{-2} g_{\mu\nu} = \kappa \braket{T_{\mu\nu}},
\end{equation}
where $\braket{T_{\mu\nu}}$ is the renormalized expectation value of the quantum stress-energy tensor (RSET) of a conformally coupled scalar field \cite{Casals:2016ioo,Casals:2016odj,Casals:2018ohr,Casals:2019jfo},
\begin{equation}
    \kappa \braket{T_{\mu\nu}(x)} = \pi l_P \lim\limits_{x' \rightarrow x} \left( 3 \nabla_\mu^x \nabla_\nu^{x'} - g_{\mu\nu} g^{\lambda\rho}\nabla_\lambda^x \nabla_\rho^{x'} - \nabla_\mu^x \nabla_\nu^{x} - \frac{1}{4 l^2} g_{\mu\nu} \right) G(x,x')\,, \;\; l_P =\frac{\hbar \kappa}{8 \pi}.
\end{equation}
Using the method of images, the propagator, $G(x,x') = \{ \phi(x),\phi(x') \}$ is the anti-commutator of the scalar field, which takes the form \cite{Steif:1993zv,Avis:1977yn,Shiraishi:1993qnr,Shiraishi:1993ti,Decanini:2005gt,Casals:2019jfo}
\begin{equation}
\label{eq:2pointfct}
    G(x, x') = \frac{1}{2 \sqrt{2} \pi} \sum\limits_{n \in I} \frac{\Theta(\sigma(x, H^n x'))}{\sqrt{\sigma(x, H^n x')}},
\end{equation}
where $\sigma(x,x')$ is the chordal distance connecting $x$ and $x'$, which can be expressed in terms of the corresponding embedding coordinates in $\mathbb{R}^{(2,2)}$ as
\begin{equation}
    \sigma(x,x') = \frac{1}{2} \left[ - \left( X^0 - X'^0 \right)^2 + \left( X^1 - X'^1 \right)^2 + \left( X^2 - X'^2 \right)^2 - \left( X^3 - X'^3 \right)^2 \right]\,.
\end{equation}
The Heaviside step function $\Theta$ in \eqref{eq:2pointfct} was introduced in \cite{Casals:2019jfo} because $\sigma(x,H^n x)$ can be negative in the rotating case. Calling $d_n(x)$ the cordal distance between a spacetime point and its $n$th image,
\begin{equation} \label{d-function}
    d_n = 2 \sigma (x, H^n x) = 2 l^2 \left[ - 1 + \cosh (2 \pi a n) \cos (2 \pi b n) - B(r) \sinh (2 \pi a n) \sin (2 \pi b n) \right],
\end{equation}
with
\begin{equation}
            B(r) = \frac{l^2 M - 2 r^2}{4 a b l^2},
\end{equation}
and the RSET takes the form \cite{Steif:1993zv,Casals:2019jfo}
\begin{equation}
\label{eq:RSET_general}
    \kappa \braket{T_{\mu\nu}} = \frac{3 l_{\mathrm{P}}}{2} \sum_{n \in I \setminus{\{0\}}} \Theta(d_n(x)) \left( S^n_{\mu\nu} - \frac{1}{3} g_{\mu\nu} g^{\lambda\rho}S^n_{\lambda\rho} \right),
\end{equation}
with
\begin{equation}
    S^n_{ab} = \frac{H^n_{ab}}{d_n^{3/2}} + \frac{3 H^n_{ac} X^c H^{-n}_{bd} X^d - H^n_{ac} X^c H^n_{bd} X^d}{d_n^{5/2}}\,. 
\end{equation}
The set $I$ in the sum \eqref{eq:RSET_general} includes all distinct images. With the splitting \eqref{splitting} between boosts ($H_a$) and rotations ($H_b)$, one must sum over different ranges for $q$ and $p$. 

\subsection{Explicit form for \texorpdfstring{$\braket{T^\mu{}_\nu}$}{T}}

Note that for any rational value of $b$ there are infinitely many values of $n$ for which $2bn$ is an integer, which occurs for $p=0$, which implies $bn=kq$ and consequently the last term in \eqref{d-function} vanishes, making the distance function $d_n$ independent of $r$. This causes an infinite number of terms in the sum \eqref{eq:RSET_general} to diverge, signaling a breakdown of the perturbative approach. This can be seen in the non-vanishing components of the stress-energy tensor,

\begin{subequations} \label{tab}
\begin{align}
\begin{split}
    \kappa \braket{T^{t}_{~\,t}} =& \frac{l_P l^2}{8 a b} \left.\sum\limits_{\substack{n=1 \\ m \nmid n}}^{\infty}\right.^\prime \left( \frac{6 \left(a^2+b^2\right) Bb_n -4 a b \bar{b}_n + 12 B \bar{a}_n}{d_n^{5/2}} \right. \\
    & \qquad\qquad \left. + \frac{\left[3 \left(a^2-b^2\right) B-2 a b\right] \left(\bar{c}_n-8\right)+\left[3 (a^2-b^2)+2abB \right] c_n e_n}{d_n^{5/2}} \right), 
\end{split}\\
    \kappa \braket{T^{t}_{~\,\theta}} =& - \frac{3 l_P l^3}{8 a b} \left.\sum\limits_{\substack{n=1 \\ m \nmid n}}^{\infty}\right.^\prime \frac{2 \left[ \left( a^2 - b^2 \right) B + 4 a b \right] b_n + 4 B a_n + \left(a^2+b^2\right) \left[B \left(\bar{c}_n - 8 \right) + e_n c_n \right]}{d_n^{5/2}}, \\
    \kappa \braket{T^{r}_{~\,r}} =& l_P \left.\sum\limits_{\substack{n=1 \\ m \nmid n}}^{\infty}\right.^\prime \frac{c_n}{d_n^{3/2}} \\
    \kappa \braket{T^{\theta}_{~\,t}} =& \frac{3 l_P l}{8 a b} \left.\sum\limits_{\substack{n=1 \\ m \nmid n}}^{\infty}\right.^\prime \frac{2 \left[ \left( a^2 - b^2 \right) B - 4 a b \right] b_n + 4 B a_n + \left(a^2+b^2\right) \left[B \left(\bar{c}_n - 8 \right) + c_n e_n \right]}{d_n^{5/2}}, \\
    \kappa \braket{T^{\theta}_{~\,\theta}} =& - \kappa \left[ \braket{T^{t}_{~\,t}} + \braket{T^{r}_{~\,r}} \right],
\end{align}
\end{subequations}
where $\sum\limits_{n}^{\quad~\prime} s_n \equiv \sum\limits_{n} \Theta(d_n) s_n$, and
\begin{subequations}
\begin{align}
a_n =& a^2 \cos (4 \pi b n) + b^2 \cosh (4 \pi a n)\;, \;\; \bar{a}_n = a^2 \cos (4 \pi b n) - b^2 \cosh (4 \pi a n), \\
b_n =& \cos (4 \pi b n) - \cosh (4 \pi a n)\;, \;\;\, \qquad  \bar{b}_n = \cos (4 \pi b n) + \cosh (4 \pi a n), \\
c_n =& 2 \cosh (2 \pi a n) \cos (2 \pi b n) + 2 \;, \;\, \, \quad \bar{c}_n = 2 \cosh (4 \pi a n) \cos (4 \pi b n) + 2, \\
e_n =& 4\sinh (2 \pi a n) \sin (2 \pi b n).
\end{align}
\end{subequations}
The presence of $B(r)$ in the numerator of the $\braket{T^{\mu}_{~\,\nu}}$ components makes them grow as $r^2$ for large distance. Hence, as the denominators are independent of $r$ for $n=qm$, these sums contain infinitely many asymptotically divergent terms. The problem is that to renormalize the stress-energy tensor using the Hadamard regularization scheme simply removes one divergent term corresponding to $n=0$ (or $p=q=0$) in the sum \eqref{eq:2pointfct}. However, we see that the stress energy tensor has infinitely many divergent terms, for $p=0$ and all possible $q$s. A ``natural'' scheme to avoid the problem would be to eliminate the $b$s that generate the issue, but this would mean eliminating all rational $b$s, contradicting \eqref{eq:b-rational}.

It is still possible in principle that, in spite of the divergences in $\braket{T^{\mu}_{~\,\nu}}$, they cancel out in the equations, yielding a finite result for the back reacted metric. We will see next that such cancellation does not occur, so that the field equations do not allow for a perturbative solution.

%% file: metric.tex
\subsection{Backreacted metric}
\label{subsec:metric}
The backreacted geometry is expected to belong in the same family of spherically symmetric stationary BTZ metrics. It is therefore natural to assume the ansatz
\begin{align}
\label{eq:metric_ansatz}
    \de s^2 =& -N(r)^2 f(r) \de t^2 + f(r)^{-1} \de r^2 + r^2 \left( \de \theta + k(r) \de t \right)^2\,.
\end{align}
Additionally, based on the previous results \cite{Casals:2019jfo} we write
\begin{align}\label{pert_ansatz-N}
    N(r) =& N_0(r) + l_P N_1(r) + O(l_P^2), \\ \label{pert_ansatz-f}
    f(r) =& f_0(r) + l_P f_1(r) + O(l_P^2), \\ \label{pert_ansatz-k}
    k(r) =& k_0(r) + l_P k_1(r) + O(l_P^2).
\end{align}
The zeroth order equations describe the unperturbed situation that yield the BTZ metric,
\begin{align}
    N_0(r) = 1, \qquad f_0(r) = \frac{r^2}{l^2} - M + \frac{J^2}{4 r^2}, \qquad k_0(r) = - \frac{J}{2 r^2}.
\end{align}
The first order corrections in $l_P$ of the field equations yield
\begin{align}
    N_1(r) =& \frac{\kappa}{l_P} \int \de r \frac{r}{f_0(r)}\left(\braket{T^{r}_{~\,r}} -  \braket{T^{t}_{~\,t}} - \frac{J}{2 r^2} \braket{T^{t}_{~\,\theta}} \right) + K_1,\\
    f_1(r) =& \int \de r \left[ -2 f_0(r) N_1'(r) + \left( \frac{J^2}{r^3} - \frac{2 M}{r} \right) N_1(r) \right. \\
    & \left. + \frac{2}{r^3} \int \de r \left( 2 M r N_1(r) + \frac{\kappa}{l_P} r^3 \braket{T^{r}_{~\,r}} \right) \right] + \frac{K_2}{r^2} + K_3,\\
    J k_1(r) =& -f_1(r) - 2 f_0(r) N_1(r) + 2 \int r \de r \left( \frac{2}{l^2} N_1(r) + \frac{\kappa}{l_P} \braket{T^{r}_{~\,r}} \right) + K_4,
\end{align}
Here the integration constants must be chosen as $K_i=0$ ($i=1,2,3,4$) so that the $O(l_P)$ metric corrections vanish for $\braket{T^{\mu}_{~\,\nu}}=0$. Even before integrating these expressions, it can be directly checked that the divergences of the stress-energy tensor do not cancel out, leading to unbounded results for $N_1$, $f_1$ and $k_1$. Consequently, the perturbative ansatz (\ref{pert_ansatz-N} --\ref{pert_ansatz-k}) does not work, since the first order corrections cannot be shown to be small.

%% file: conclusions.tex
\section{Summary} \label{sec:concl}

We have shown that a naked singularity of an overspinning BTZ geometry conformally coupled to a quantum scalar field does not lead to a renormalized stress-energy tensor. This causes incurable infinities to appear in the equations of motion and in the purportedly perturbative solutions. This is contrary to the previously studied cases of conical singularities, where the quantum corrections of the conformally coupled scalar field yields a finite renormalized stress-energy tensor and the resulting back-reacted geometry acquires a horizon, which provides a mechanism that enforces cosmic censorship
\cite{Casals:2016ioo, Casals:2016odj, Casals:2018ohr, Casals:2019jfo}. Our result indicates that the overspinning geometry is plagued by a more severe form of naked singularity, inaccessible by a perturbative approach. Consequently, it is not possible to claim that the singularity may become dressed by perturbative quantum corrections.

Our result seems to indicate that coupling a conformal quantum scalar field to an overspinning geometry may cause the metric to be significantly different from the original BTZ metric. In any event, it is not possible to assert, as in the other cases of naked singularities, that quantum mechanics provides a cosmic censor in this case. 

It would be interesting to understand whether there is a more profound problem with this type of geometry, or if the strongly rotating behavior simply prevents the application of perturbative methods. Perhaps one way to approach this problem would be by numerical methods, hoping to get a better understanding of the nature of this particular type of singularity and to see if this is purely a problem of the perturbative approach, or if there is a more fundamental issue with the overspinning singularity.